\documentclass[prd,twocolumn,tightenlines,showpacs,nofootinbib]{revtex4-1} %% Ben's package
\usepackage{hyperref} 
\hypersetup{pdfborder={0 0 0}}  %Removes awful bright hyper-link borders
\usepackage{xcolor}
\definecolor{blue}{rgb}{0.,0.,0.5}   % a more subtle than default blue
\hypersetup{
	colorlinks,
	linkcolor={blue},
	citecolor={blue},
	urlcolor={blue}
}

\usepackage{amsfonts}
\usepackage{dcolumn}
\usepackage{bm}
\usepackage{amsmath}
\usepackage{nicefrac}
\usepackage{amsthm}
\usepackage{amssymb}
\usepackage{graphicx}
\usepackage{color}
\newcommand{\appropto}{\mathrel{\vcenter{
  \offinterlineskip\halign{\hfil$##$\cr
    \propto\cr\noalign{\kern2pt}\sim\cr\noalign{\kern-2pt}}}}}

\newcommand{\bra}[1]{\langle #1|}	%Dirac Bras
\newcommand{\ket}[1]{|#1\rangle}	%Dirac Kets
\renewcommand{\v}[1]{\boldsymbol{#1}}		%bold-math for vectors
\newcommand{\g}{\gamma} %for gamma matrices. 
\newcommand{\matr}[4]{\begin{pmatrix}#1&#2\\#3&#4\end{pmatrix}}	% 2x2 matrix

\definecolor{blue2}{rgb}{0.,0.,1} 
\newcommand{\removed}[1]{{\color{red} XXX}}                              %anything I removed is placed here. It won't be compiled, but should be romved before resubmission, to avoid confusion to the editors.
\begin{document}

%%%%%%%%%%%%%%%%%%%%%%%%%%%%%%%%%%%%%%%%%%%%%%%%%%%
%\title{Constraints on $\mathcal{CPT}$-odd and Lorentz invariance-violating physics from anomalous magnetic dipole moments of the electron and muon}
\title{Tests of CPT and Lorentz symmetry from muon anomalous magnetic dipole moment %%from\\ anomalous magnetic dipole moments of the electron, proton and muon
}

\date{\today}
\author{Y.~V.~Stadnik} \email[]{y.stadnik@unsw.edu.au}
\author{B.~M.~Roberts} \email[]{b.roberts@unsw.edu.au}
\author{V.~V.~Flambaum}
\affiliation{School of Physics, University of New South Wales, Sydney 2052, Australia}

%%%%%%%%%
\begin{abstract}
%We derive the relativistic factor for splitting of the $g$-factors of a fermion and its anti-fermion partner, which is important for placing constraints on dimension-5, CPT-odd and Lorentz invariance-violating interactions from experiments performed in a cyclotron. From existing data, we extract limits (1$\sigma$) on the coupling strengths of a background field (including the field amplitude), which is responsible for such $g$-factor splitting, with an electron, proton and muon to be: $\left| f^0_e \right| < 1.4 \times 10^{-25} \textrm{~GeV~T}^{-1}$, $\left| f^0_p \right| < 2 \times 10^{-22} \textrm{~GeV~T}^{-1}$ and $\left| f^0_\mu \right|  < 5 \times 10^{-24} \textrm{~GeV~T}^{-1}$, respectively, in the laboratory frame. From existing data, we also extract limits (1$\sigma$) on the coupling strengths of related dimension-5 interactions of a background field with an electron, proton, neutron and muon: $| \v{d}_e^{(\perp)} | < 10^{-22} \textrm{~GeV~T}^{-1}$, $| \v{d}_p^{(\perp)} | < 10^{-23} \textrm{~GeV~T}^{-1}$, $| \v{d}_n^{(\perp)} | < 10^{-22} \textrm{~GeV~T}^{-1}$ and $| \v{d}_\mu^{(\perp)} | < 10^{-22} \textrm{~GeV~T}^{-1}$, respectively, in the laboratory frame.

We derive the relativistic factor for splitting of the $g$-factors of a fermion and its anti-fermion partner, which is important for placing constraints on dimension-5, $CPT$-odd and Lorentz-invariance-violating interactions from experiments performed in a cyclotron. 
From existing data, we extract limits (1$\sigma$) on the coupling strengths of the temporal component, $f^0$, of a background field (including the field amplitude), which is responsible for such $g$-factor splitting, with an electron, proton, and muon: 
$|f^0_e|< 2.3 \times 10^{-12} ~\mu_{\textrm{B}}$, 
$|f^0_p|< 4 \times 10^{-9} ~\mu_{\textrm{B}}$, 
and 
$|f^0_\mu|< 8 \times 10^{-11} ~\mu_{\textrm{B}}$, 
respectively, in the laboratory frame. 
From existing data, we also extract limits on the coupling strengths of the spatial components, $d^{\perp}$,  of related dimension-5 interactions of a background field with an
 electron, proton, neutron, and muon: 
$| {d}_e^{\perp} | \lesssim 10^{-9} ~\mu_{\textrm{B}}$, 
$| {d}_p^{\perp} | \lesssim 10^{-9} ~\mu_{\textrm{B}}$, 
$| {d}_n^{\perp} | \lesssim 10^{-10} ~\mu_{\textrm{B}}$, 
and 
$| {d}_\mu^{\perp} | \lesssim 10^{-9} ~\mu_{\textrm{B}}$, 
respectively, in the laboratory frame.

\end{abstract}

\pacs{11.30.Er, 11.30.Cp, 14.60.Ef, 14.60.Cd} %%  CPT-invariance (Parity, CP, also), Lorentz invariance, properties of muons, properties of electrons   

\maketitle 

%%%%%%%%%%
\section{Introduction}
%%\label{Sec:Introduction}
%%\emph{Introduction.} --- 
The violation of the fundamental symmetries of nature is an area of substantial interest, both experimentally and theoretically. Field theories, which are constructed from the principles of locality, spin-statistics and Lorentz invariance, conserve the combined $\mathcal{CPT}$ symmetry. The violation of one or more of these three principles, presumably from some form of ultra-short distance scale physics, opens the door for the possibility of $\mathcal{CPT}$-odd physics. $\mathcal{CPT}$-odd and Lorentz invariance-violating physics has been sought for experimentally in the form of the coupling 
\begin{equation}
\label{conv_Delta-E}
\hat{H}_{\textrm{int}} = \v{b}\cdot\v{\sigma} 
\end{equation}
between a background cosmic field, $\v{b}$, and the spin of an electron, proton, neutron and muon, $\v{\sigma}$, and constraints on the strengths of such interactions have been obtained \cite{Berglund1995,Bluhm2000a,Hou2003,Cane2004,Heckel2006,*Heckel2008,Bennett2008CPT,Altarev2009,Gemmel2010,Brown2010,Peck2012,Allmendinger2014}. For further details on the broad range of experiments performed in this field and a brief history of the improvements in these limits, we refer the reader to the reviews of \cite{Kostelecky1999,Kostelecky2011data} and the references therein. Electric dipole moment (EDM) measurements have also been proposed as sensitive probes of $\mathcal{CPT}$-odd physics \cite{Bolokhov2008}. Limits on $\mathcal{P}$-odd fermion effects induced by $\mathcal{CPT}$-odd, Lorentz invariance-violating couplings have been extracted from existing parity nonconservation (PNC) and anapole moment data \cite{Roberts2014}. Atomic dysprosium has been proposed for odd-parity tests of Lorentz symmetry \cite{Leefer2013odd} and has been used to place limits on local Lorentz invariance \cite{Hohensee2013limits}. Pion and kaon systems have also been suggested for tests of Lorentz invariance \cite{Picek1983_LNI}.

Dirac theory predicts that all elementary standard model (SM) fermions should have the gyromagnetic ratio $g^{\textrm{Dirac}} = 2$. Quantum field theory corrections result in deviations from $g^{\textrm{Dirac}} = 2$, which can be quantified by the anomalous magnetic dipole moment (MDM) parameter
\begin{equation}
\label{AMM_general}
a = \frac{g-2}{2} .
\end{equation}
Consider, for instance, the particularly interesting case of the muon. The current SM prediction for the anomalous MDM of the muon consists of quantum electrodynamic, weak and hadronic contributions \cite{Beringer2012PDG} (see also the multitude of references therein for more details of some of the pioneering theory and experiments, which led to the current refined prediction of $a_\mu^{\textrm{SM}}$):
\begin{equation}
\label{AMM_muon_SM}
a_\mu^{\textrm{SM}} = 116591803(1)(42)(26) \times 10^{-11} ,
\end{equation}
where the uncertainties are due to the electroweak, lowest-order hadronic and higher-order hadronic contributions, respectively. The most accurate measurement to date for the anomalous MDMs of the muon and anti-muon are \cite{Mohr2012codata,Bennett2002,*Bennett2004,*Bennett2006final}:
\begin{eqnarray}
\label{AMM_muon_exp-}
&& a_\mu^{\textrm{exp}} = 116592150(80)(30) \times 10^{-11} ,\\		
\label{AMM_muon_exp+}
&& a_{\bar{\mu}}^{\textrm{exp}} = 116592040(60)(50) \times 10^{-11} ,			
\end{eqnarray}
respectively, and, assuming $\mathcal{CPT}$-invariance and taking into account correlations between systematic uncertainties, gives the average
\begin{equation}
\label{AMM_muon_exp}
a_\mu^{\textrm{exp}} = 116592091(54)(33) \times 10^{-11} ,			
\end{equation}
where the quoted uncertainties are due to statistical and systematic sources, respectively. The result (\ref{AMM_muon_exp}) represents about an order-of-magnitude improvement in precision compared with the now classic experiment of \cite{Bailey1979final}. The difference between the SM prediction and experimental value:
\begin{equation}
\label{AMM_muon_diff}
\Delta a_\mu = a_\mu^{\textrm{exp}} - a_\mu^{\textrm{SM}} = 288(80) \times 10^{-11} ,
\end{equation}
with experimental and theoretical uncertainties added in quadrature, represents a discrepancy between the two values of 3.6 $\sigma$, suggesting that the effects of new physics beyond the SM may be manifesting themselves. Some of the most promising current explanations for this discrepancy are supersymmetric models \cite{Barbieri1982,Ellis1982,Francis1991,Moroi1996muon,Carena1997,Czarnecki2001muon,Feng2001supersymmetry,Baltz2001,Baer2001,Martin2001,Heinemeyer2004,Stockinger2007muon,Marchetti2009tan} and the dark photon, which is a massive vector boson from the dark matter sector that couples to SM particles by mixing with the ordinary photon \cite{Fayet2007u,Pospelov2009_DP,Tucker2011muonic}. See also Refs.~\cite{Miller2007muon,Jegerlehner2009muon,Miller2012muon,Queiroz2014} for some of the more recent reviews on the muon anamalous MDM puzzle. Future measurements of the anomalous MDM of the muon with increased precision are currently planned \cite{Muon-new_FermiLab}.

Tests of $\mathcal{CPT}$-odd and Lorentz invariance-violating physics from measurements of the anomalous MDMs of various particles have been proposed previously \cite{Bluhm1997,Bluhm1998,Bluhm2000,Deile2001-cpt_muon,Bolokhov2005}. In the framework of the 
Kosteleck\'y \emph{et al.}
Standard Model Extension (SME) parametrisation \cite{Bluhm1997,Bluhm1998,Bluhm2000}, it was found that there were no leading-order corrections to the $g$-factors of fermions and their respective anti-fermions. Instead, measurements of magnetic field-independent splittings of the anomalous precession frequencies for a fermion and its respective anti-fermion were proposed for placing limits on $\mathcal{CPT}$-odd and Lorentz invariance-violating physics. In the dimension-5 framework of \cite{Bolokhov2005}, it was demonstrated that the $g$-factors of an electron and a positron may be split by a $\mathcal{CPT}$-odd and Lorentz invariance-violating interaction, and a limit on the relevant interaction parameter was extracted from existing data at the time. 

In the present work, we consider $\mathcal{CPT}$-odd, Lorentz invariance-violating dimension-5 couplings, which are linear in the gauge field strength. We derive the relativistic factor for splitting of the $g$-factors of a fermion and its anti-fermion partner, which is important for placing constraints on dimension-5, $\mathcal{CPT}$-odd and Lorentz invariance-violating interactions from experiments performed in a cyclotron. Anomalous MDMs are ideal physical quantities for tests of $\mathcal{CPT}$-odd and Lorentz invariance-violating physics, because of the high precision with which these quantities can be determined. We extract limits on the coupling strengths of the background field, which splits $g$-factors, with an electron, proton and muon. We also extract limits on the coupling strengths of related dimension-5 interactions of a background field with an electron, proton, neutron and muon. For details of other most recently proposed laboratory tests of $\mathcal{CPT}$-odd and Lorentz invariance-violating physics with muons, we refer the reader to Ref.~\cite{Kostelecky2014_muons}.

%%%%%%%%%%
\section{Theory}
%%\label{Sec:Theory}
%%\emph{Theory.} --- %% 
%%{\color{red} STATIC CF!!!}
We employ the natural units $\hbar = c =1$ and the metric signature $(+---)$ in this work. Various $\mathcal{CPT}$-odd couplings of dimension-5 have been classified (see e.g.~\cite{Myers2003Ultraviolet,Nibbelink2005Lorentz,Bolokhov2005,Bolokhov2008}). In the present work, we consider the following $\mathcal{CPT}$-odd, Lorentz invariance-violating dimension-5 couplings, which are linear in the gauge field strength \cite{Bolokhov2008} (see also Ref.~\cite{Bolokhov2005} for a more detailed discussion of some of the relevant terms):
\begin{align}
\label{dim-5_CPT}
\mathcal{L} = - \sum & \left[ c^\nu \bar{\psi}_f \gamma^\lambda F_{\lambda \nu} \psi_f + d^\nu \bar{\psi}_f \gamma^\lambda \gamma^5 F_{\lambda \nu} \psi_f           \right. \notag \\
&\left. + f^\nu \bar{\psi}_f \gamma^\lambda \gamma^5 \tilde{F}_{\lambda \nu} \psi_f + g^\nu \bar{\psi}_f \gamma^\lambda \tilde{F}_{\lambda \nu} \psi_f  \right] ,
\end{align}
where the sum is over all SM fermions $f$ and SM gauge groups, with 
{$F_{\lambda \nu}$ and $\tilde{F}_{\lambda \nu}$} representing the field and dual field tensor strengths, respectively. 
The terms 
$c^\nu$, $d^\nu$, $f^\nu$ and $g^\nu$ in (\ref{dim-5_CPT}) represent the amplitudes of the background cosmic field(s) with the corresponding interaction strength amalgamated into them. In the present work, we are interested in systems exposed to external magnetic and electric fields. From the Lagrangian (\ref{dim-5_CPT}), we find the following interaction Hamiltonians:
\begin{eqnarray}
\label{f0-couple}
&&\hat{H}_{\textrm{int}}^A = f^0 \v{B} \cdot \v{\Sigma} ,\\
%----
\label{d-couple}
&&\hat{H}_{\textrm{int}}^B = (\v{d} \times \v{B}) \cdot \v{\Sigma} ,\\
%-----
\label{f-couple}
&&\hat{H}_{\textrm{int}}^C = - \v{f} \cdot \v{B} \gamma^5 ,\\
\label{c-couple}
&&\hat{H}_{\textrm{int}}^D = (\v{c} \times \v{B}) \cdot \v{\alpha} ,\\
\label{g0-couple}
&&\hat{H}_{\textrm{int}}^E = g^0 \v{B} \cdot \v{\alpha} ,\\
\label{g-couple}
&&\hat{H}_{\textrm{int}}^F = -\v{g} \cdot \v{B} , \\
%------------------------------------------------------------------------------------------------
\label{d0-couple}
&&\hat{H}_{\textrm{int}}^G = d^0 \v{E} \cdot \v{\Sigma} ,\\
%----
\label{f2-couple}
&&\hat{H}_{\textrm{int}}^H = -(\v{f} \times \v{E}) \cdot \v{\Sigma} ,\\
%-----
\label{d2-couple}
&&\hat{H}_{\textrm{int}}^I = - \v{d} \cdot \v{E} \gamma^5 ,\\
\label{g2-couple}
&&\hat{H}_{\textrm{int}}^J = -(\v{g} \times \v{E}) \cdot \v{\alpha} ,\\
\label{c0-couple}
&&\hat{H}_{\textrm{int}}^K = c^0 \v{E} \cdot \v{\alpha} ,\\
\label{c2-couple}
&&\hat{H}_{\textrm{int}}^L = -\v{c} \cdot \v{E} , 
\end{eqnarray}
where $\v{B}$ is the external magnetic field strength, $\v{E}$ is the external electric field strength; 
$\v{\Sigma} \equiv \matr{\v{\sigma}}{0}{0}{\v{\sigma}}$,
$\v{\alpha} \equiv \g^0\v{\g}$,
and $\gamma^5 \equiv i \gamma^0 \gamma^1 \gamma^2 \gamma^3$ are Dirac matrices, and we have suppressed the possible dependence of the cosmic field parameters on the fermion species $f$ in our notation.

We are specifically interested in interactions that can alter the spin-precession frequency of a fermion to leading-order. Of the interactions (\ref{f0-couple}) to (\ref{c2-couple}), only (\ref{f0-couple}) alters the spin-precession frequency of a fermion to leading-order (in fact the interaction (\ref{f0-couple}) splits the $g$-factors of fermions and their corresponding anti-fermions), since the other interactions satisfy at least one of the following three criteria:

(I) The interaction produces no observable effect;

(II) The interaction mixes opposite-parity states;

(III) The interaction produces sidereal effects, which average to zero after the averaging of acquired data.

Crucial to the point (I) above are the identity \cite{Ginges_PhysRep2004}:
\begin{equation}
\label{id-Ginges}
\v{E} \cdot \v{\Sigma} = \frac{[\v{\Sigma} \cdot \v{\nabla},\hat{H}]}{e} ,
\end{equation}
where $\hat{H}$ is the full Dirac Hamiltonian and $-e$ is the electron charge, and the more general identity
\begin{equation}
\label{id-new}
E_j \Sigma_k = \frac{[\Sigma_k \nabla_j,\hat{H}]}{e} .
\end{equation}
Note that none of the interactions (\ref{f0-couple}) to (\ref{c2-couple}) affect the cyclotron frequency.

Regarding possible $\mathcal{P}$-odd interactions, we mention that the interactions (\ref{c-couple}) and (\ref{g0-couple}) do not contribute to PNC amplitudes \cite{Ginges_PhysRep2004} of the form:
\begin{align}
\label{PNC_amp}
E_{\rm PNC}^{a \to b} = \sum_n \left[ \frac{\bra{b}\v{\hat{d}}\ket{n}\bra{n} \hat{H}_{\textrm{int}} \ket{a}}{E_a-E_n} +\frac{\bra{b} \hat{H}_{\textrm{int}} \ket{n}\bra{n} \v{\hat{d}} \ket{a}}{E_b-E_n} \right]
\end{align}
for transitions $\ket{a} \to \ket{b}$, where $\hat{H}_{\textrm{int}}$ is a perturbing $\mathcal{P}$-odd operator and $\v{\hat{d}}$ is the electric dipole ($E1$) operator, because of the identity $\v{\alpha} = i [\hat{H}, \v{r} ]$. Also, matrix elements of (\ref{c-couple}) and (\ref{g0-couple}) between a pair of nearly degenerate levels of opposite parity are negligibly small. Interaction (\ref{f-couple}) can give rise to sidereal non-zero PNC amplitudes and matrix elements between opposite parity levels, which do not necessarily scale as the energy difference between the pair of opposite parity levels. PNC amplitudes of this nature are determined entirely by relativistic effects and such matrix elements are typically dominated by relativistic effects (see e.g.~Refs.~\cite{Stadnik2014,Roberts2014}).

\section{Results}
\emph{Electrons and protons.---}  
We first consider the interaction with electrons and protons described by (\ref{f0-couple}) in the non-relativistic limit, which is the appropriate description of experiments for determining anomalous MDMs using a Penning trap, where the Lorentz factor is $\gamma = (1-v^2)^{-1/2} \approx 1$. 
In an applied magnetic field, the combined potentials experienced by an electron and positron due to the interactions of their MDMs with the magnetic field and also their interactions via (\ref{f0-couple}) may be written as
\begin{eqnarray}
\label{Ue-}
&& U_e = \left(\frac{g_e \mu_{\textrm{B}}}{2} + f^0_e \right) \v{\sigma}_e \cdot \v{B} , \\
\label{Ue+}
&& U_{\bar{e}} = \left(-\frac{g_e \mu_{\textrm{B}}}{2} + f^0_e \right) \v{\sigma}_{\bar{e}} \cdot \v{B} ,
\end{eqnarray}
%%In the non-relativistic limit, Eq.~(\ref{f0-couple}) can be rewritten in the equivalent form as {\color{red} --- Maybe just include a sigma dot B with magnetic terms explicitly (i.e. potential energies!!!!) --- and so on here... for e-, p+, e+, and p- separately, so there's no confusion...?? 4 equations here, but not for muon, to make it absolutely clear!!! }
respectively, where $\mu_{\textrm{B}}$ is the Bohr magneton. Likewise, for a proton and anti-proton, the potentials may be written as
\begin{eqnarray}
\label{Up+}
&& U_p = \left(-\frac{g_p \mu_{\textrm{N}}}{2} + f^0_p \right) \v{\sigma}_p \cdot \v{B} , \\
\label{Up-}
&& U_{\bar{p}} = \left(\frac{g_p \mu_{\textrm{N}}}{2} + f^0_p \right) \v{\sigma}_{\bar{p}} \cdot \v{B} ,
\end{eqnarray}
respectively, where $\mu_{\textrm{N}}$ is the nuclear magneton.
The resulting splitting in the $g$-factors of an electron and a positron can thus be expressed as
\begin{equation}
\label{e-e+}
a_{e}^{\textrm{exp}} - a_{\bar{e}}^{\textrm{exp}} = 2a_{e}^{\textrm{CF}} ,
\end{equation} 
where $a_e^{\textrm{CF}} = f_e^0 / \mu_{\textrm{B}}$, whereas the splitting in the $g$-factors of a proton and anti-proton can thus be expressed as
\begin{equation}
\label{p+p-}
a_{\bar{p}}^{\textrm{exp}} - a_{p}^{\textrm{exp}} = 2a_{p}^{\textrm{CF}} ,
\end{equation} 
where $a_p^{\textrm{CF}} = f_p^0 / \mu_{\textrm{N}}$. 

In the absence of a recent experimental value for $a_{\bar{e}}$, we extract a limit on $f_e^0$ from the SM prediction and experimentally measured values for $a_{e}$. This is likely to be a weaker limit than that, which may be extracted from $a_e$ and a future value for $a_{\bar{e}}$, since the new physics (such as supersymmetry), which contribute equally to both $a_e$ and $a_{\bar{e}}$, is likely to occur at a lower energy scale than that for $\mathcal{CPT}$-violating physics, which may result in a splitting of $a_e$ and $a_{\bar{e}}$. This is already borne out, for instance, in the muon anomalous MDM values in (\ref{AMM_muon_exp-}), (\ref{AMM_muon_exp+}) and (\ref{AMM_muon_diff}). Noting that the SM prediction for the anomalous MDM of the electron is \cite{de2013update} (see also Refs.~\cite{Schwinger_term_a-e_1949,Petermann57A,*Petermann58B,Somm57A,Laporta1996,Aoyama2007a,Aoyama2012,Bouchendira2011,VanDyck1987,Odom2006} for some of the pioneering theory and experiments, which led to the current refined prediction of $a_e^{\textrm{SM}}$):
\begin{equation}
\label{AMM_electron_SM}
a_e^{\textrm{SM}} = 1159652181.82(78) \times 10^{-12} ,
\end{equation}
with associated uncertainties added in quadrature, and that the experimentally measured value for the anomalous MDM of the electron is
\cite{Hanneke2008,Mohr2012codata}:
\begin{equation}
\label{AMM_electron_exp}
a_e^{\textrm{exp}} =  1159652180.76(27) \times 10^{-12} ,
\end{equation}
we extract the limit ($1 \sigma$) on the coupling strength of the background field with an electron via interaction (\ref{f0-couple}) to be
\begin{equation}
\label{f0e-limit}
|f^0_e|< 2.3 \times 10^{-12} ~\mu_{\textrm{B}} ,
\end{equation}
in the laboratory frame. Further measurements, in particular of the positron anomalous MDM, with increased precision would lead to a more stringent constraint on $f^0_e$.

The most accurate measurement to date for the MDM of the proton is \cite{Proton_MDM_2014-exp}:
\begin{equation}
\label{MDM_proton_exp}
\mu_p^{\textrm{exp}} =  2.792847350(9) ~ \mu_{\textrm{N}} ,
\end{equation}
while for the anti-proton \cite{DiSciacca2013}:
\begin{equation}
\label{MDM_proton_exp}
\mu_{\bar{p}}^{\textrm{exp}} =  -2.792845(12) ~ \mu_{\textrm{N}} , 
\end{equation}
from which we extract the limit ($1 \sigma$) on the coupling strength of the background field with a proton via interaction (\ref{f0-couple}) to be
\begin{equation}
\label{f0p-limit}
|f^0_p|< 4 \times 10^{-9} ~\mu_{\textrm{B}} ,
\end{equation}
in the laboratory frame.

\emph{Muons.---}  
We now consider the interaction 
with muons
described by (\ref{f0-couple}) in the relativistic case, which is the appropriate description of experiments for determining $a_\mu$ using a cyclotron, where $\gamma \gg 1$. In this case, Eq.~(\ref{f0-couple}) can be expressed in the same form for a muon and anti-muon:
\begin{equation}
\label{f0-couple2_rel}
\hat{H}_{\textrm{int}}^A = f^0_\mu \left[ \v{B} \cdot \v{\sigma}_\mu + \frac{2(\v{p}_\mu \cdot \v{\sigma}_\mu)(\v{p}_\mu \cdot \v{B}) - p_\mu^2 (\v{B} \cdot \v{\sigma}_\mu) }{(\gamma+1)^2 m_\mu^2} \right] ,
\end{equation}
where $\v{p}_\mu$ is the muon relativistic momentum and $m_\mu$ is the muon mass. In a cyclotron, $\v{p}_\mu$ and $\v{B}$ are perpendicular and so (\ref{f0-couple2_rel}) simplifies to
%\begin{equation}
%\label{f0-couple2_rel_2}
%\hat{H}_{\textrm{int}} = f_0^\mu \v{B} \cdot \v{\sigma} \left[1 - \frac{\gamma^2 v^2}{(\gamma+1)^2} \right] .
%\end{equation}
\begin{equation}
\label{f0-couple2_rel_2}
\hat{H}_{\textrm{int}}^A = f^0_\mu \v{B} \cdot \v{\sigma}_\mu \left[1 - \frac{\gamma^2 v_\mu^2}{(\gamma+1)^2} \right] .
\end{equation}
The background field contribution to the observed anomalous MDM of the muon is
\begin{equation}
\label{muon_a-cf}
a_\mu^{\rm CF} = \frac{2 f^0_\mu m_\mu}{e}\left[1 - \frac{\gamma^2 v^2_\mu}{(\gamma+1)^2} \right] .
\end{equation}
Note that in the ultra-relativistic limit ($\gamma \to \infty$), the correction to the anomalous MDM of the muon from (\ref{muon_a-cf}) vanishes. In the experiment of \cite{Bennett2006final}, $\gamma = 29.3$ and so there is only a finite suppression of the contribution to the anomalous MDM of the muon arising from (\ref{f0-couple}). 
%Noting that the muon magnetic dipole moment operator may be written as 
%\begin{equation}
%\label{muon_MDM}
%\v{\mu}_{\mu} = -\frac{\mu_{\textrm{B}} m_e \v{\sigma}}{m_\mu} \left[\frac{g_\mu}{2} + \frac{1-\gamma}{\gamma} \right] ,
%\end{equation}
The anomalous precession frequency of a muon or anti-muon in a cyclotron can be written, with account of both SM and cosmic field contributions, as
\begin{equation}
\label{omega_a-muon}
\v{\omega}_a = \frac{e}{m_\mu} \left[\tilde{a}_\mu \v{B} - \left(\tilde{a}_\mu - \frac{1}{\gamma^2 - 1} \right) \v{v}_\mu \times \v{E} \right]  ,
\end{equation}
where $\tilde{a}_\mu = a_\mu^{\textrm{SM}} + a_\mu^{\textrm{CF}}$ for a muon and $\tilde{a}_\mu = a_\mu^{\textrm{SM}} - a_\mu^{\textrm{CF}}$ for an anti-muon, with $a_\mu^{\textrm{CF}}$ given by (\ref{muon_a-cf}). The experimentally chosen Lorentz factor $\gamma = 29.3$ ensures that the $\v{v}_\mu \times \v{E}$ term in (\ref{omega_a-muon}) is significantly suppressed compared to the first term. The splitting in the $g$-factors of a muon and an anti-muon in this case can be expressed as
\begin{equation}
\label{e-e+}
a_{\mu}^{\textrm{exp}} - a_{\bar{\mu}}^{\textrm{exp}} = 2a_{\mu}^{\textrm{CF}} .
\end{equation} 
From the values in (\ref{AMM_muon_exp-}) and (\ref{AMM_muon_exp+}), we extract the limit ($1 \sigma$) on the coupling strength of the background field with a muon via interaction (\ref{f0-couple}) to be
\begin{equation}
\label{f0mu-result}
|f^0_\mu|< 8 \times 10^{-11} ~\mu_{\textrm{B}} ,
\end{equation}
in the laboratory frame.

\section{Other tests}
%%\emph{Other tests.} --- 
It was pointed out in Ref.~\cite{Bolokhov2008} that EDMs may serve as sensitive tests of $\mathcal{CPT}$-odd interactions, in association with the $d^0$ term in (\ref{dim-5_CPT}). Here we also mention a further test stemming from the interaction Hamiltonian (\ref{d-couple}), which in the non-relativistic limit reads
\begin{equation}
\label{d-couple_NRL}
\hat{H}_{\textrm{int}}^B = (\v{d} \times \v{B}) \cdot \v{\sigma} .
\end{equation}
The interaction (\ref{d-couple_NRL}), like (\ref{conv_Delta-E}), produces sidereal shifts of the energy levels in atomic and nuclear systems, but only in the presence of an external magnetic field, and can be sought for in a similar manner to the coupling (\ref{conv_Delta-E}) via sidereal modulations of transition frequencies \cite{Berglund1995,Bluhm2000a,Hou2003,Cane2004,Heckel2006,*Heckel2008,Bennett2008CPT,Altarev2009,Gemmel2010,Brown2010,Peck2012,Allmendinger2014}. %Note that if the same background cosmic field gives rise to both of the interactions (\ref{conv_Delta-E}) and (\ref{d-couple_NRL}), then there will be a phase shift of $\pi / 2$ radians between the induced energy shifts for the two interactions. 
Existing experiments were performed in non-zero magnetic fields and so we can extract limits on $| {d}^{\perp} |$ from the magnetic field strengths used in these experiments and existing limits on $\left| \v{b} \right|$ (Table \ref{tab:limits_d}). Here $| {d}^{\perp} |$ is the magnitude of the largest component of $\v{d}$, which is perpendicular to the applied magnetic field, at any time during the experiment.
 %% and can seek both using same apparatus...

Finally, we mention that the effects discussed in this work do not need to be restricted solely to static cosmic fields. Dynamic cosmic fields (one particularly important example of which is axion dark matter) are also possible and analogous effects in an oscillatory form, with oscillation frequencies determined by the mass of the field excitation, may be induced. For further details, see, e.g.~Refs.~\cite{Graham2011,*Graham2013,Budker2013C,Stadnik2014,Roberts2014}. A network of domain wall-type structures of cosmic fields are also be possible and these may induce transient effects analogous to those discussed in this work. For further details, see, e.g.~Refs.~\cite{Pospelov2013,Pospelov-Derev14,Stadnik2014defects}.

%%%%
  \begin{table}[h!]
    \centering%
    \caption{Limits ($1\sigma$) on the interaction strengths of a background cosmic field with an electron, proton, neutron and muon via interaction (\ref{d-couple}). Limits are derived for $| {d}^{\perp} |$ using existing experimental limits on $\left| \v{b} \right|$ and the magnetic field strengths employed in these experiments. $X$ denotes fermion species.} 
\begin{ruledtabular}%
\begin{tabular}{cccccc}
 & \multicolumn{3}{c}{Experiment}    &  \multicolumn{1}{c}{This work}  \\ 
\cline{2-4}\cline{5-5}
\multicolumn{1}{c}{$X$}   & \multicolumn{1}{c}{$\left| \v{b}_X \right|$ / GeV}   & \multicolumn{1}{c}{$\left| \v{B} \right|$ / T} & \multicolumn{1}{c}{Ref.}  & \multicolumn{1}{c}{$| {d}^{\perp}_X |$ / $\mu_B$}\\ 
\hline
$e$  & $10^{-29}$  & $10^{-7}$ & \cite{Heckel2006,*Heckel2008}  & $10^{-9}$  \\ 
$p$ & $10^{-28}$  & $5.5 \times 10^{-7}$ & \cite{Peck2012}  & $10^{-9}$  \\ 
$n$ & $10^{-29}$  & $5.5 \times 10^{-7}$ & \cite{Peck2012}  & $10^{-10}$  \\
$\mu$ & $10^{-23}$  & $1.45$ & \cite{Bennett2008CPT} & $10^{-9}$  \\
  \end{tabular}%
\end{ruledtabular}%  
    \label{tab:limits_d}%
  \end{table}%.
%%%%%%%%%%

%%%%%%%%
\section*{ACKNOWLEDGEMENTS}
%%\emph{Acknowledgements.} ---
We would like to thank Maxim Pospelov and Alan Kosteleck\'y for useful discussions. This work is supported in part by the Australian Research Council and by the Perimeter Institute for Theoretical Physics. Research at the Perimeter Institute is supported by the Government of Canada through Industry Canada and by the Province of Ontario through the Ministry of Economic Development \& Innovation.

%%%%%%%%%%%%%%%%%%%%%%%%%%%%%%%%%%%%%%% 
%\begin{thebibliography}{99} 

%\bibliographystyle{unsrtnat}
%\bibliographystyle{aipauth4-1}
%\bibliographystyle{apsrev}
%\bibliographystyle{abbrv}
%\bibliography{Muon_AMM}

\bibliography{Muon_AMM}

\end{document}